\documentclass[twocolumn]{aastex631}
\usepackage{sidecap}
\usepackage{wrapfig}
\usepackage{float}
\usepackage{xcolor}

\accepted{AJ - 31 October 2022}

\shorttitle{KELT-20 Dayside}
\shortauthors{Kasper et al.}
\graphicspath{{./}}

\begin{document}
\title{Unifying High- and Low-resolution Observations to Constrain the Dayside Atmosphere of KELT-20b/MASCARA-2b}
\correspondingauthor{David Kasper}
\email{kasperd@uchicago.edu}

\author[0000-0003-0534-6388]{David Kasper}
\affil{Department of Astronomy \& Astrophysics, University of Chicago, 5640 South Ellis Avenue, Chicago, IL 60637, USA}

\author[0000-0003-4733-6532]{Jacob L.\ Bean}
\affil{Department of Astronomy \& Astrophysics, University of Chicago, 5640 South Ellis Avenue, Chicago, IL 60637, USA}

\author[0000-0002-2338-476X]{Michael R.\ Line}
\affil{School of Earth and Space Exploration, Arizona State University, Tempe, AZ 85281, USA}

\author[0000-0003-4526-3747]{Andreas Seifahrt}
\affil{Department of Astronomy \& Astrophysics, University of Chicago, 5640 South Ellis Avenue, Chicago, IL 60637, USA}

\author[0000-0003-2404-2427]{Madison T. Brady}
\affil{Department of Astronomy \& Astrophysics, University of Chicago, 5640 South Ellis Avenue, Chicago, IL 60637, USA}

\author[0000-0003-3667-8633]{Joshua Lothringer}
\affil{Department of Physics, Utah Valley University, Orem, UT 84058, USA}

\author[0000-0002-1321-8856]{Lorenzo Pino}
\affil{INAF-Osservatorio Astrofisico di Arcetri Largo Enrico Fermi 5I-50125 Firenze, Italy}

\author[0000-0002-3263-2251]{Guangwei Fu}
\affil{Department of Astronomy, University of Maryland, College Park, MD 20742, USA}

\author[0000-0002-8573-805X]{Stefan Pelletier}
\affil{Institut de Recherche sur les Exoplan\`etes, D\'epartement de Physique, Universit\'e de Montr\'eal, 1375 Avenue Th\'er\`ese-Lavoie-Roux, Montreal, H2V 0B3, Canada}

\author[0000-0002-4410-4712]{Julian St\"urmer}
\affil{Landessternwarte, Zentrum f\"ur Astronomie der Universit\"at Heidelberg, K\"onigstuhl 12, 69117 Heidelberg, Germany}

\author[0000-0001-5578-1498]{Bj\"orn Benneke}
\affil{Institut de Recherche sur les Exoplan\`etes, D\'epartement de Physique, Universit\'e de Montr\'eal, 1375 Avenue Th\'er\`ese-Lavoie-Roux, Montreal, H2V 0B3, Canada}

\author[0000-0002-7704-0153]{Matteo Brogi}
\affil{Department of Physics, University of Warwick, Coventry CV4 7AL, UK}
\affil{INAF-Osservatorio Astrofisico di Torino, Via Osservatorio 20,
I-10025 Pino Torinese, Italy}
\affil{Centre for Exoplanets and Habitability, University of Warwick, Coventry, CV4 7AL, UK}

\author{Jean-Michel D\'esert}
\affil{Anton Pannekoek Institute for Astronomy, University of Amsterdam, 1098 XH Amsterdam, The Netherlands}

\begin{abstract}

We present high-resolution dayside thermal emission observations of the exoplanet KELT-20b/MASCARA-2b using the MAROON-X spectrograph. Applying the cross-correlation method with both empirical and theoretical masks and a retrieval analysis, we confirm previous detections of Fe\,\textsc{i} emission lines and we detect Ni\,\textsc{i} for the first time in the planet (at 4.7$\sigma$ confidence). We do not see evidence for additional species in the MAROON-X data, including notably predicted thermal inversion agents TiO and VO, their atomic constituents Ti\,\textsc{i} and V\,\textsc{i}, and previously claimed species Fe\,\textsc{ii} and Cr\,\textsc{i}. We also perform a joint retrieval with existing \textit{Hubble Space Telescope}/WFC3 spectroscopy and \textit{Spitzer}/IRAC photometry. This allows us to place bounded constraints on the abundances of Fe\,\textsc{i}, H$_2$O, and CO, and to place a stringent upper limit on the TiO abundance. The results are consistent with KELT-20b having a  solar to slightly super-solar composition atmosphere in terms of the bulk metal enrichment, and the carbon-to-oxygen and iron-to-oxygen ratios. However, the TiO volume mixing ratio upper limit (10$^{-7.6}$ at 99\% confidence) is inconsistent with this picture, which, along with the non-detection of Ti\,\textsc{i}, points to sequestration of Ti species, possibly due to nightside condensation. The lack of TiO but the presence of a large H$_2$O emission feature in the WFC3 data is challenging to reconcile within the context of 1D self-consistent, radiative-convective models.

\end{abstract}


\section{Introduction} \label{sec:intro}

\begin{deluxetable*}{lccccccccclcc}
\tabletypesize{\scriptsize}
\tablecolumns{13}
\tablewidth{0pc}
\tablecaption{\label{tab:obs_log} Log of the MAROON-X observations}
\tablehead{
 \colhead{UT Date}  & \colhead{Exposures} &  \colhead{Planet Phase Range} & \colhead{Airmass} & \colhead{Conditions} & \colhead{Seeing} & \colhead{Average Blue Arm SNR}
}
\startdata
2021 May 29 10:47 $\rightarrow$ 14:03 & 34 & 0.63 $\rightarrow$ 0.67 & 1.28 $\rightarrow$  1.03 & stable, clear & 0.6\arcsec &  176\\ 
2021 June 04 11:37 $\rightarrow$ 14:34 & 31 & 0.37 $\rightarrow$ 0.40 & 1.09 $\rightarrow$ 1.02 $\rightarrow$ 1.08 & stable, clear & 0.5\arcsec & 196\\
\enddata
\end{deluxetable*}

KELT-20b/MASCARA-2b (hereafter KELT-20b for brevity) is a 1.56\,$R_{\text{Jup}}$ radius exoplanet orbiting a bright ($m_V \sim 7.6$) A star with a 3.47\,day orbital period \citep{lund17,Talens18}. These parameters place it in the population of the $\sim30$\footnote{\url{https://exoplanetarchive.ipac.caltech.edu}} known transiting hot Jupiters orbiting early-type stars. KELT-20b is very highly for dayside emission atmospheric follow-up due to the apparent magnitude of the host star, the size of the planet relative to the host star, and the planet's high equilibrium temperature of $\sim$2250\,K. Furthermore, the unambiguous detection of both iron \citep{borsa22,Yan22,Johnson22} and water \citep{Fu22} in the emission spectrum of this exoplanet, which is one of the coolest exoplanets with a detected thermal inversion, gives a special opportunity to unify ground-based, high-resolution and space-based, low-resolution observations of exoplanet atmospheres.

The thermal emission spectrum of KELT-20b has been previously observed with ground-based, high-resolution spectrographs by \citet{borsa22}, \citet{Yan22}, and \citet{Johnson22}. Using data from HARPS-N, \citet{borsa22} found a 6.8$\sigma$ detection of Fe\,\textsc{i}, as well as 3$\sigma$ detections of Fe\,\textsc{ii} and Cr\,\textsc{i}. The \citet{borsa22} detections of Fe\,\textsc{ii} and Cr\,\textsc{i} were only obtained for their post-secondary eclipse dataset and did not appear in their pre-eclipse dataset, thus suggesting the presence of chemical inhomogeneities. \citet{Yan22} used data from CARMENES and found Fe\,\textsc{i} at 7.5$\sigma$. Using PEPSI, \citet{Johnson22} found Fe\,\textsc{i} in emission at 15.1$\sigma$ with non-detections of Fe\,\textsc{ii} and Cr\,\textsc{i}, among others. All the detected species in the previous works were observed in emission, which indicates a thermal inversion in KELT-20b's atmosphere. \citet{Johnson22} inferred upper limits on the abundances of their TiO, VO, FeH, and CaH non-detections with volume mixing ratios in the $\sim 10^{-9}-10^{-10}$ range, with the exception of FeH, for which they found at an upper limit of $3\times10^{-7}$.

With space-based data, \citet{Fu22} found H$_{2}$O and CO by analyzing the emission spectrum of KELT-20b derived from \textit{Hubble Space Telescope}/WFC3 and \textit{Spitzer}/IRAC eclipse observations. The water feature is the strongest of those in emission observed to-date with WFC3 \citep{Mansfield2021}. \citet{Fu22} claimed that TiO is necessary to cause the deep inversion that leads to the large water emission feature. However, this requirement is in tension with the previous ground-based measurements that all failed to see TiO in KELT-20b's thermal emission spectrum.

In this paper we present ground-based, high-resolution observations of the dayside emission from KELT-20b that were obtained using the MAROON-X instrument on the Gemini North telescope \citep{seifahrt16, seifahrt18, seifahrt20}. We aim to determine the composition of the planet's dayside and perform a joint analysis including existing \textit{HST} and \textit{Spitzer} data to unify ground- and space-based results for this planet. We present our observations in \S\ref{sec:obs}. We describe two analyses of the data in \S\ref{sec:template} and \ref{sec:ret} that yield the detection of the planet's atmosphere and constraints on its properties. We contextualize our atmospheric retrieval with a comparison to atmospheric forward models in \S\ref{sec:RCEmodels}. We conclude with a discussion of the results in \S\ref{sec:discussion}.

\section{Observations} \label{sec:obs}
We used MAROON-X to obtain 4.51\,hours of data (6.2\,hours total observing time including overheads) of KELT-20b during a post-secondary eclipse phase on May 29 2021 and a pre-secondary eclipse phase on June 04 2021 UTC (Program ID: GN-2021A-Q119). We obtained data with both the MAROON-X ``blue'' (500 -- 663\,nm) and ``red'' (654 -- 920\,nm) channels. The two channels were observed simultaneously with 250\,s exposures. Table~\ref{tab:obs_log} gives a log of the observations.

The data were reduced with the standard MAROON-X pipeline \citep{seifahrt20}. The pipeline includes detector calibration, one-dimensional spectral extraction, barycentric corrections calculated for the flux-weighted midpoint of each observation, and wavelength solutions and instrumental drift corrections based on the simultaneous Fabry-Perot etalon calibration data.

\section{Detection of the Planetary Atmosphere Signal} \label{sec:template}
We performed two analyses of our data to characterize KELT-20b's dayside atmosphere. The first analysis used a cross-correlation function (CCF) technique with line-weighted binary templates following \citet{pino20} and \citet{kasper2021}. We used our established approach to create a time (i.e., phase) evolving spectrum containing planet lines normalized to the planet plus star continuum. We used eleven stellar templates built from observed stellar spectra \citep[e.g.,][]{suarez20} as masks for the CCF analysis. These masks correspond to F9, G2, G8, G9, K2, K6, M0, M2, M3, M4, and M5 stellar types and come from the ESPRESSO data reduction pipeline (``ESPRESSO DRS''\footnote{\url{http://eso.org/sci/software/pipelines/}}). The mask ensemble spans the transition from ionized and neutral atoms to molecules as absorption features in stellar spectra. We used the masks in emission as analogs for the dayside of the hydrogen-dominated atmosphere of KELT-20b. 

\begin{figure}[t!]
\begin{center}
\includegraphics[width=\linewidth]{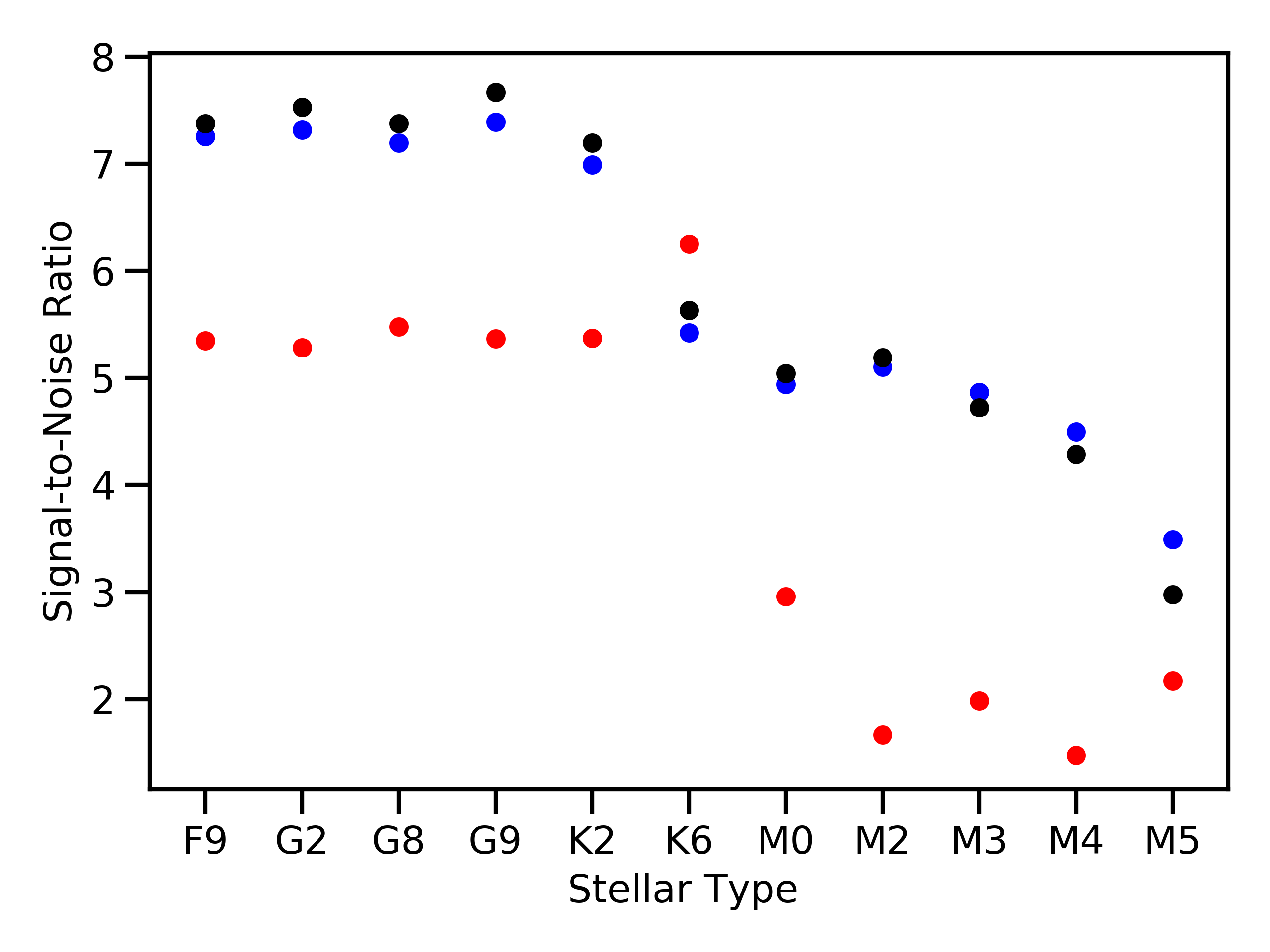}
\caption{Detection SNR of the planetary atmospheric signal as a function of template stellar type. The responses of the red and blue channels are tracked individually by corresponding color. The combination of the two channels is shown in black.}
\label{fig:ccf_comp}
\end{center}
\end{figure}

Figure \ref{fig:ccf_comp} shows the peak signal-to-noise ratio (SNR) from the CCF analysis when combining the pre- and post-eclipse datasets and utilizing each of the masks. The SNR normalization was computed via a 3$\sigma$ background clipping method \citep[as in][]{kasper2021}.
The high significance correlation with the earlier spectral type masks implicates neutral atomic lines, and Fe\,\textsc{i} in particular, as the dominant opacity source in KELT-20b's atmosphere at the wavelengths probed. In comparison with the earlier-type masks applied to the blue channel, M-type masks give lower detection confidences, down to a non-detection at M5. Additionally, M-type masks do not yield a significant detection in the red channel. This indicates, in agreement with previous ground-based results, that molecules like TiO are likely not present in the planet's atmosphere.

\begin{figure}
\begin{center}
\includegraphics[width=\linewidth]{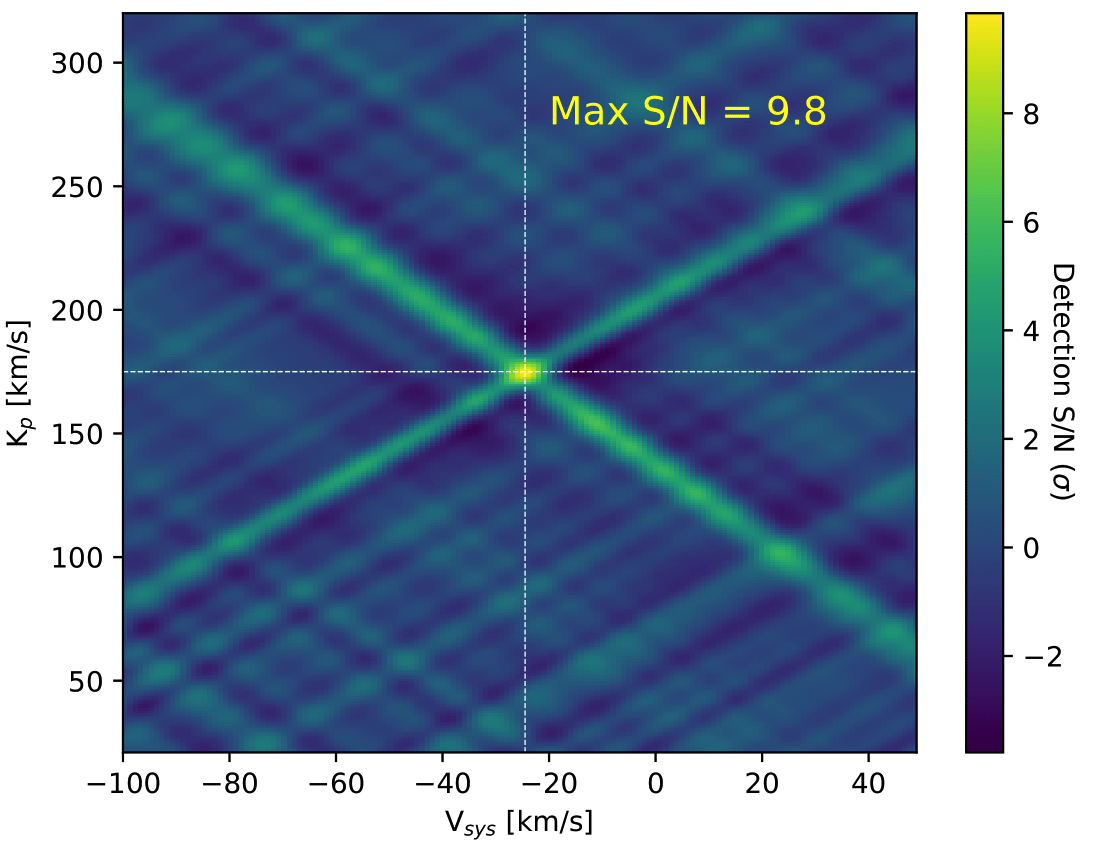}
\caption{Map of the CCF in the planet velocity ($K_{p}$) vs.\ system velocity ($V_{\text{sys}}$) plane using the theoretical Fe\,\textsc{i} mask. Dotted lines correspond to the known system velocity and exoplanet orbital speed of -24.5\,km\,s$^{-1}$ and 175\,km\,s$^{-1}$, respectively, as in \citet{borsa22}.}
\label{fig:FeI}
\end{center}
\end{figure}

Complementary to the above analysis, we also explored the initial detectability of a variety of specific gases via data-model template cross-correlation \citep{kasper2021,line2021}. As in past works we removed stationary telluric and stellar features via the singular value decomposition method (we removed between 2-8 singular values and found little difference. We ultimately settled on 2 to prevent accidental removal of the planetary signal). The high-resolution template models are derived from the converged output of a KELT-20b-specific 1D radiative-convective-thermochemical equilibrium (1D-RC) model calculated using the {\tt ScCHIMERA} code \citep{pis18,Arcangeli2018,Mansfield2021, kasper2021} assuming solar composition. The high-resolution model spectra are then convolved with a planetary rotation kernel and a Gaussian spectral response function at an R\,=\,85,000, as appropriate for MAROON-X. We searched for CaH, Ca\,\textsc{i} and \textsc{ii}, CrH, Cr\,\textsc{i}, FeH, Fe\,\textsc{i} and \textsc{ii}, MgH, Mg\,\textsc{i}, Na\,\textsc{i}, Ni\,\textsc{i}, Si\,\textsc{i}, Ti\,\textsc{i} and \textsc{ii}, TiO, V\,\textsc{i} and \textsc{ii}, and VO.

Of all the species tested only Fe\,\textsc{i} and Ni \textsc{i} were found with a significant response (9.8$\sigma$ and 4.7$\sigma$, respectively). We also explored a grid of constant-with-altitude abundances for each gas \citep[using the same converged temperature-pressure profile found above; e.g.,][]{Giacobbe2021}  to ensure we did not miss any species detections due to the assumption of thermochemical equilibrium. No additional species were found in this analysis.

Figure~\ref{fig:FeI} shows an SNR map for Fe\,\textsc{i} in the planet velocity (K$_{p}$) vs.\ system velocity (V$_{\text{sys}}$) plane (a similar plot for Ni\,\textsc{i} can be found in the Appendix). The SNR in the map was computed via a 3-sigma background clipping method \citep[as in][]{kasper2021}. With the combination of pre- and post-eclipse phases we find agreement with the system velocity and planetary orbital speed. This $K_{p}$ vs.\ $V_{\text{sys}}$ mapping was also performed on the stellar template CCFs to ensure that the signal originated from the planet.

\begin{deluxetable}{lcl}
\tablecaption{\label{tab:priors} Description of retrieved parameters and their prior ranges. All priors are assumed uniform between the bounds given. Variables correspond to the labeling in the corner plot shown in the Appendix.}
\tablehead{
 \colhead{Parameter}  & \colhead{Description} &  \colhead{Prior} 
}
\startdata
$K_{p\text{,pre/post}}$ & Keplerian velocity for  &  150 -- 210\,km\,s$^{-1}$\\
                 &  pre/post eclipse nights & \\
$V_{\text{sys,pre/post}}$ & systemic velocity  & -50 -- 10\,km\,s$^{-1}$ \\
log($a_{\text{pre/post}}$) & model multiplicative  & -1 -- 1\\
& scale factor & \\
log($\gamma_1$) & vis-to-IR opacity & -3 -- 4 \\ 
log($\kappa_{\text{IR}}$) & IR opacity & -3 -- 0 (cgs) \\
T$_{\text{irr}}$ & irradiation temperature & 1000 -- 5000\,K\\
$\Delta$log\,$g$ & differential log-gravity & -1 -- 0 (cgs)\\
& from max (log\,$g$=3.28) & \\
H$^-$, Fe, TiO, & log gas volume & -12 -- 0 \\
 H$_2$O, CO & mixing ratios & \\
H*e$^-$  & log mixing ratio& -18 -- 0 \\
& hydrogen $\times$ electron & \\
& for free-free cont. & \\
\enddata
\end{deluxetable}

\section{Retrieval Analysis} \label{sec:ret}
Following \citet{line2021} and \citet{kasper2021}, we applied the \citet{Brogi2019} cross-correlation-to-log-likelihood retrieval framework to derive abundances and the vertical temperature structure in KELT-20b's dayside atmosphere. The \texttt{CHIMERA} forward model underlying the retrieval assumed constant-with-altitude abundances and used the \citet{Guillot2010} parameterization of the temperature-pressure (T-P) profile. The retrieval parameters and their prior ranges are given in Table \ref{tab:priors}. A more detailed description of the radiative transfer method, including opacity sources, and log-likelihood implementation is given in \cite{line2021} and \cite{kasper2021} (with opacity sources there-in including \cite{mckemmish19,John1988,Grimm2015,Grimm2021}) as well as new cross-sections utilized in this work from the EXOPLINES data base \citep{GharibNezhad21} as sourced from the line lists for VO \citep{Mckemmish16}, H$_2$O \citep{Polyansky2018}, MgH \citep{GharibNezhad2013}, CrH and CaH \citep{Bernath2020MOLLIST}.

For our ``MAROON-X only'' retrieval we combined the blue and red arm MAROON-X data for both the pre- and post-eclipse observations. We also performed a joint retrieval on the combination of the MAROON-X observations and the \textit{HST}/WFC3 and \textit{Spitzer}/IRAC 4.5\,$\mu$m observations presented in \citet{Fu22}. The joint retrieval was performed by combining the high-resolution and low-resolution likelihoods as described in \cite{Brogi2019}.  In all the retrievals we included H$_2$O, CO, TiO, Fe\,\textsc{i}, and abundance proxies for the H$^-$ bound-free and free-free continua. The detection of Ni\,\textsc{i} occurred after we performed the retrieval analysis. Because the mass of the planet is not well constrained \citep[upper limit of 3.5\,M$_{\text{Jup}}$,][]{lund17}, we also included a log\,$g$ free parameter that spanned a range between maximum log\,$g$ (3.28, cgs) and one dex below. The results are summarized in Figure \ref{fig:retrieval} and the full corner plot is given in the Appendix (Figure~\ref{fig:staircase}). We note that in comparing the maximum log-likelihood difference between the two retrievals we found the difference as entirely due to the additional $red\chi^2 \sim\,1$ data-points and that the MAROON-X contribution did not change in any meaningful way. This is unsurprising as there is little overlap in common parameters between the two datasets, with the exception of the T-P profile. E.g., water and CO do not present themselves in the MAROON-X data and the atomic species don’t show up in the WFC3. As relevant below, this suggests that any T-P profile information arising from the low-res contribution is fully consistent with what the MAROON-X data prefers.

\begin{figure*}
\begin{center}
\includegraphics[width=\textwidth]{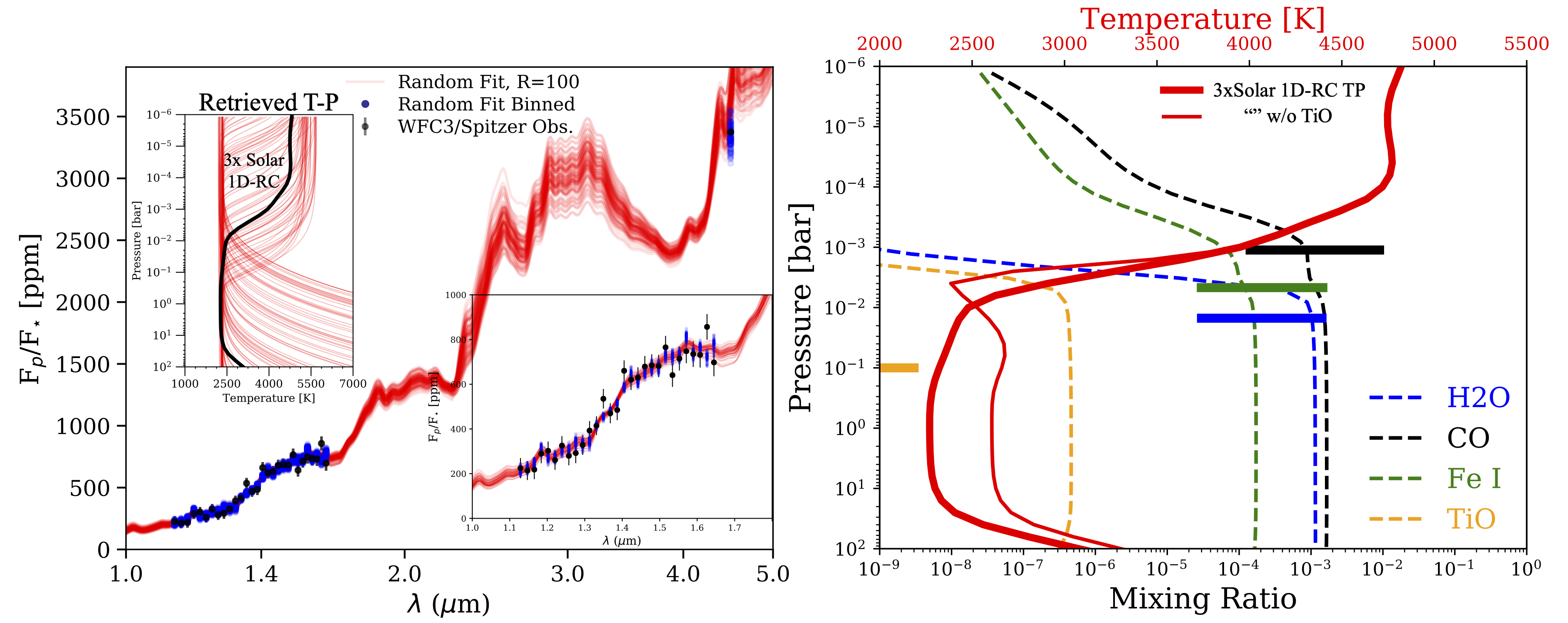}
\caption{Summary of the joint retrieval analysis. The \textit{left} panel shows the model fits to the space-based data resulting from the retrieval on the MAROON-X, \textit{HST}/WFC3, and \textit{Spitzer}/IRAC data. The red curves represent the spectra produced via 100 random draws from the posterior (smoothed to an R $=100$). The blue points are those same spectra binned to the WFC3 wavelengths. The smearing in the blue points illustrates the spread in the binned model spectra. The inset in the bottom right of the left panel shows a zoom in on the WFC3 data. The ``Retrieved T-P" inset in the upper left shows reconstructed T-P profiles from 100 posterior draws from the combined MAROON-X, WFC3, and \textit{Spitzer} (blue) observations. A $3\times$Solar 1D-RC model T-P profile from \texttt{ScCHIMERA} is shown in black for comparison. The \textit{right} panel summarizes the retrieved abundance constraints in the context of a $3\times$solar 1D-RC model. The solid horizontal bars give the retrieved 68\% confidence intervals on the chemical species. The position of the bars on the y axis (pressure axis) are at the approximate location probed by the wavelengths sensitive to those species (see Figure \ref{fig:tau_sfc} in the appendix). The predicted thermochemical mixing ratio profiles are shown as dashed lines and the temperature profile is shown as the thick red curve (governed by the top x axis). The thinner red temperature profile is the same $3\times$solar scenario but with the opacity due to TiO/VO removed. }
\label{fig:retrieval}
\end{center}
\end{figure*}

\begin{figure*}
\begin{center}
\includegraphics[width=\textwidth]{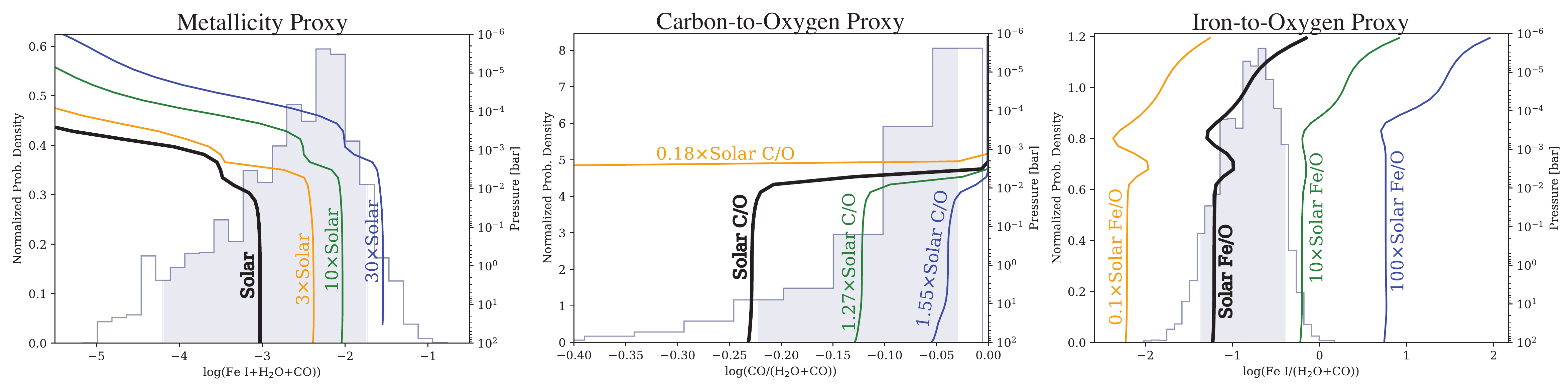}
\caption{Comparison of retrieved elemental abundance proxies to those predicted from a battery of 1D-RC models. In each panel the histogram (scale on the left y-axis) is derived via a Monte Carlo propagation of the retrieved samples for Fe,\textsc{i}, H$_2$O, and CO into the quantity given under each x-axis ({\it left}:~log(Fe\,\textsc{i}+H$_2$O+CO)), {\it middle}:~log(CO/(H$_2$O+CO))), {\it right}:~log(Fe\,\textsc{i}/(H$_2$O+CO))). The light shaded region is the 95\% confidence interval for each proxy.  The solid curves are the vertical abundance proxy profiles (pressure scale on the right y-axis of each panel) derived from the indicated 1D-RC models. In each panel, all elemental abundances are scaled uniformly by the indicated value relative to solar. The retrieved abundance proxy constraints are consistent with solar-composition, under the 1D-RC assumption.  }
\label{fig:abund_summary}
\end{center}
\end{figure*}

The joint retrieval to all the datasets provides an excellent fit to the \textit{HST} and \textit{Spitzer} observations (median $\chi^2$ per number of data points is $\sim 0.9$, p-value=0.02).  As expected from the results in the literature and the analysis in \S\ref{sec:template}, the retrieval prefers a strong thermal inversion.  The temperature gradient proxy parameter, log($\gamma_1$), is +21$\sigma$ away from zero, where a value of zero indicates an isothermal T-P profile, negative is a monotonically decreasing T-P profile, and positive is an inverted T-P profile (increasing temperature with decreasing pressure above the inversion level). The combined MAROON-X + \textit{HST} and \textit{Spitzer} retrieval provides a much more stringent constraint on this temperature gradient proxy than with MAROON-X alone (only 1.8$\sigma$ above zero, see blue vs. red histograms as well as additional caption details in Figure~\ref{fig:staircase}). We note that the {\it pressure level} at which the primary line forming temperature gradient resides is not well constrained [controlled by the log($\kappa_{IR}$) parameter] due to the lack of prior constraint on gravity and the degeneracy between metallicity and the ``$\tau$=2/3" pressure level. In the top left inset on the left panel of Figure~\ref{fig:retrieval}, this lack of constraint is apparent as the reconstructed T-P profiles invert at a continuum of pressure levels, with pressures below the inverted portion naturally becoming isothermal due to the \citet{Guillot2010} profile parameterization employed, as it assumes a double gray formalism. The retrieved T-P profile {\it gradient} (as fully apparent in some reconstructions within the bounds of the figure) matches quite well the expectation from 1D-RC. However, this forward model includes TiO, which is seemingly absent given the data. We explore this issue more in the next section (\S\ref{sec:RCEmodels}).

The joint retrieval between the optical and near-to-mid IR enables constraints on both refractory (Fe) and volatile (C, O) elements. We derived bounded constraints on the abundances of Fe\,\textsc{i}, H$_2$O, and CO (Figure~\ref{fig:staircase}). The retrieved abundances are found to be in qualitative agreement with the expectations for a $\sim$3$\times$solar composition gas in thermochemical equilibrium (see the right panel of Figure~\ref{fig:retrieval}). Inferring elemental abundance ratios from retrieved molecular abundances, especially with simplistic retrieval forward modeling assumptions, (e.g., constant-with-altitude volume mixing ratios) is not always straightforward [e.g., \cite{Sheppard2017} vs. \cite{Arcangeli2018}]. This is especially true in ultra-hot Jupiters where thermal dissociation can deplete the abundances of measurable species (e.g., Fe\,\textsc{i}, H$_2$O) into non-measured species (Fe\,\textsc{ii}, OH, O\,\textsc{i}).  In order to place the retrieved abundances into context, we compare a series of elemental abundance proxies to a battery of, again, 1D-RC models, summarized in Figure~\ref{fig:abund_summary}. 

We use the total sum of the bounded retrieved gases (Fe\,\textsc{i}+H$_2$O+CO) as a proxy for total metal enrichment (a.k.a, ``metallicity"), the ratio, CO/(H$_2$O+CO) as a tracer of the carbon-to-oxygen ratio (C/O), and Fe\,\textsc{i}/(H$_2$O+CO) as a proxy for the iron-to-oxygen ratio (Fe/O). Figure~\ref{fig:abund_summary} shows these secondary retrieval data-products as histograms compared to those same quantities (and their dependencies with altitude) from different composition 1D-RC models. We find that the overall metallicity (left most panel) is consistent with enrichment values $\lessapprox 30\times$solar, with a relatively loose lower bound of $\sim$0.1$\times$solar. Depending on the exact pressure level probed, the C/O (middle panel) can range anywhere between solar (C/O=0.55) to super-solar (C/O=0.85). Finally, we find the most stringent constraint on the Fe/O (right most panel).  The retrieved values are largely consistent with solar, if not up to a few times solar (but below 10$\times$). 

We can be more quantitative about the elemental abundance constraints if we make further assumptions about the pressure levels probed to correct for potential biases due to the constant-with-altitude gas mixing ratio assumption. As discussed above, we are unable to constrain the absolute pressure-level location of the base of the inversion. However, these degeneracies work out in such away that the $\tau$=2/3 level probed by the different wavelengths should always see the same temperature gradient, and largely the same abundance along a mixing ratio profile. An example of the pressure levels probed within the 3$\times$solar 1D-RC is shown in the Appendix ($\sim$0.1\,bar to 0.1\,mbar, see Figure~\ref{fig:tau_sfc}).

Additionally, the constant-with-altitude gas mixing ratios assumed in the retrieval will be more heavily weighted towards the deeper, greater abundance layers, before molecular dissociation and atomic ionization occurs. Conveniently, our chosen elemental abundance proxies are fairly constant with pressure/altitude (within a 1D-RC) at layers deeper than the dissociation/ionization level (as also can be seen by the pressure levels probed in Figure~\ref{fig:tau_sfc}). It is also true that in these deeper layers the C/O and Fe/O proxies are {\it exact} measures for those elemental ratios as Fe\,\textsc{i}, H$_2$O, and CO are the sole carriers of elemental Fe, O, and C (at altitudes where dissociation begins, O becomes predominately locked into OH and O\,\textsc{i}, in addition to CO, and Fe into Fe\,\textsc{ii}).

Given the above caveats and assumptions, we provide log of the abundances relative to the uniform regions of the solar composition 1D-RC model (thick black curves in each Figure \ref{fig:abund_summary} panel): [Fe\,\textsc{i}+H$_2$O+CO] = -1.28 -- 1.49, [CO/(H$_2$O+CO)]=-0.47 -- 0.19, and  [Fe\,\textsc{i}/(H$_2$O+CO)] = -0.20 -- 0.86 at 95\% confidence (here ``[X/Y]'' is the usual bracket notation used for stellar abundances where zero is equal to solar). Taken together, the elemental abundance ratios are relatively unremarkable and largely consistent with solar-composition (perhaps a modest enhancement in the Fe/O and overall metallicity), and generally in-line with current trends \citep[e.g.][]{kreidberg14,Benneke2015} -- any interpretation beyond that would not be supported by these data-sets. 

The retrieval also provides an upper limit on the TiO volume mixing ratio of 10$^{-7.6}$ at 99\% confidence. This is about one order of magnitude lower than the expectation from 3$\times$solar 1D-RC model. It has long been recognized that TiO could potentially play a major role in causing thermal inversions in hot Jupiter atmospheres \citep[e.g.,][]{Hubeny2003, Fortney2008}. However, detections of this species have been rare and not without controversy \citep[see][and references therein]{prinoth22}. KELT-20b has a strong thermal inversion, yet we and others have failed to find TiO in the emission spectrum \citep{Yan22, borsa22, Johnson22}.

\begin{figure*}
\begin{center}
\includegraphics[width=\linewidth]{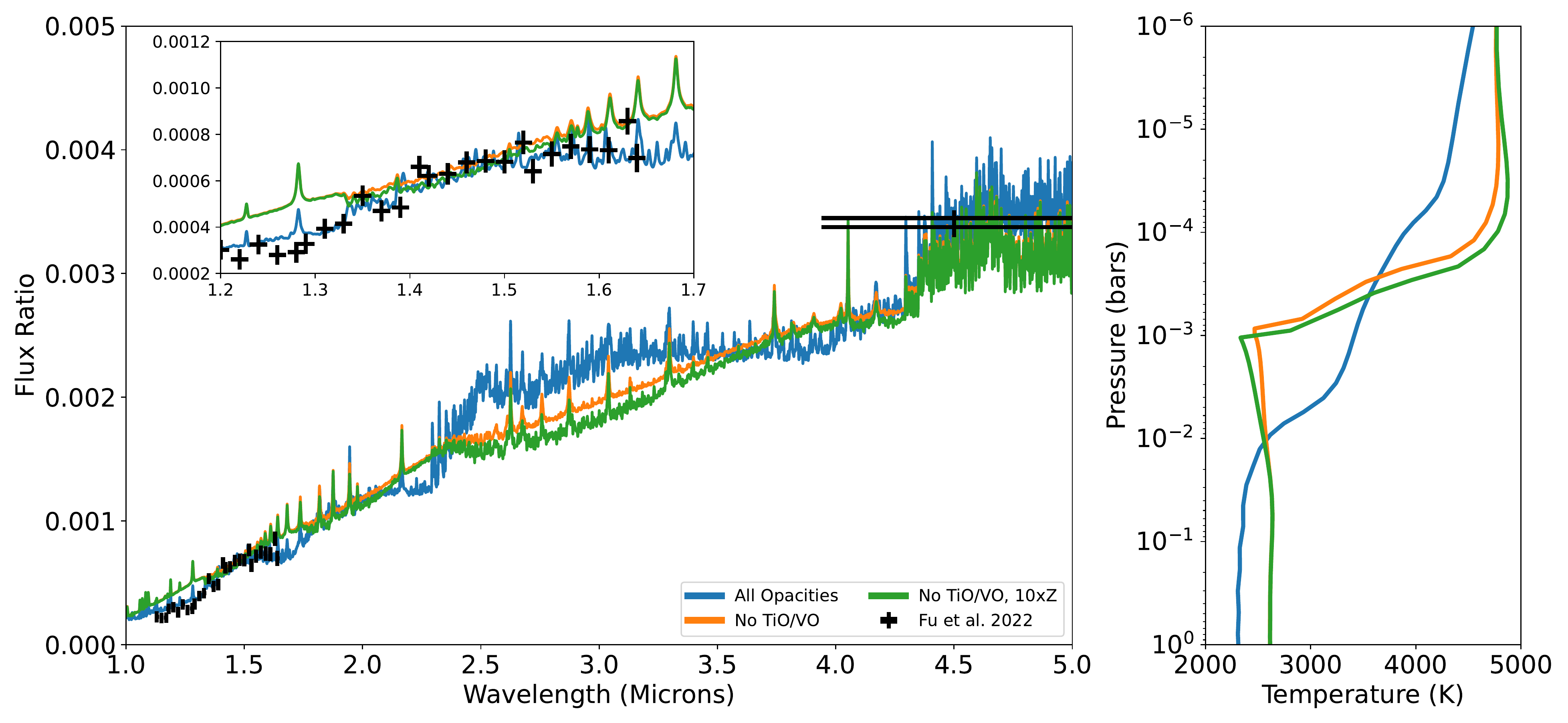}
\caption{Secondary eclipse spectra (planet-to-star flux ratio, \textit{left}) and T-P profiles (\textit{right}) for KELT-20b from \texttt{PHOENIX} model calculations (lines) compared to the \textit{HST} and \textit{Spitzer} data of \citet[][points]{Fu22}. The inset zooms in on the \textit{HST}/WFC3 data. The nominal model assuming thermochemical equilibrium for a solar composition gas is shown in blue. Models with TiO and VO opacity artificially removed are shown in orange (solar metallicity) and green ($10\times$ solar metallicity). The primary impact of this opacity is its influence on the thermal structure of the atmosphere rather than specific spectral features in the plotted region. The sharp emission lines in the spectra are a result of absorption lines expected in the stellar spectrum.}
\label{fig:phoenix}
\end{center}
\end{figure*}

\section{Radiative-Convective Equilibrium Atmosphere Models} \label{sec:RCEmodels}
The lack of TiO in the atmosphere of KELT-20b raises the question of what causes the strong thermal inversion. To explore this issue we modeled a number of scenarios with the \texttt{PHOENIX} 1D self-consistent radiative-convective equilibrium atmosphere model \citep{hauschildt:1999,Barman2001}. Our ultra-hot Jupiter model set-up was similar to those presented in \cite{lothringer18}. We first modeled the atmosphere of KELT-20b with all available opacities, including atomic opacity from hydrogen up to uranium, continuous opacity sources like H$^{-}$, and molecular opacity, including TiO and VO. The resulting spectrum shows a qualitative match to the low-resolution \textit{HST} and \textit{Spitzer} observations of \citet{Fu22}, in agreement with the \texttt{ScCHIMERA} self-consistent models. However, we also computed a model without TiO and VO, since we do not find evidence of these species with our high-resolution observations. The results are summarized in Figure~\ref{fig:phoenix}.

While the \texttt{PHOENIX} models with TiO and VO opacity artificially removed are a better match to the results of our (and others') high-resolution observations, importantly, they do not match the low-resolution, space-based observations. This is due to the fact that without TiO and VO, the heating from the absorption of the short-wavelength stellar irradiation by species like the atomic metals is balanced by cooling by molecules like H$_{2}$O and CO. An inversion only forms at lower pressures once these latter molecules have thermally dissociated and can no longer radiatively cool the atmosphere. This means that the region that H$_{2}$O probes in the thermal emission spectrum is below the large temperature inversion and we would not expect to see H$_{2}$O in emission (see Figure \ref{fig:retrieval}). On the other hand, when TiO and VO are present, they can heat up the atmosphere in the region that H$_2$O probes, resulting in a strong H$_2$O emission feature. Thus, the non-detection of TiO and VO at high-resolution is in tension with the strong H$_2$O emission feature found by \cite{Fu22}.

\citet{lothringer19} pointed out the importance of the host star spectral energy distribution in setting the thermal structure of highly irradiated planets. \citet{Fu22} suggested that KELT-20b's A-type host star, with its higher proportion of short-wavelength flux than the typical hot Jupiter host star, is likely responsible for the large water emission feature seen in the planet's spectrum. Therefore, if TiO is not present to cause the thermal inversion at the deep pressures where H$_2$O is still intact then it could be that the models are missing opacity at the short wavelengths where the host star is particularly bright.

\section{Discussion} \label{sec:discussion}
Our detection of atomic emission lines agrees with the broad conclusions of recent publications analyzing the dayside of the ultra-Hot Jupiter KELT-20b: there is a thermal inversion in the atmosphere and neutral iron (as opposed to TiO or VO) is the dominant optical opacity source observed in the spectrum \citep{borsa22,Yan22,Johnson22}. Additionally, for the first time with KELT-20b we search for and find Ni\,\textsc{i} (at 4.7$\sigma$ confidence). 

KELT-20b is one of the coolest exoplanets known to have a thermal inversion and it has the strongest water emission feature seen in over two dozen hot Jupiter emission spectra obtained with \textit{HST}/WFC3 \citep{Fu22}. However, there have been no detections of TiO or atomic Ti in either its emission spectrum or in the many observations of its transmission spectrum \citep{belloarufe22, Langeveld22, Rainer21, nugroho20, Stangret20, hoeijmakers20, Kesseli20,  Casasayas19}. Given the presence of the other refractory species that have been detected for this planet (in addition to those described already, Na, Mg, Ca, and Cr species have also been detected in the transmission spectrum), sequestration of Ti in the unobservable parts of the atmosphere due to cold trapping seems a likely explanation \citep{Spiegel2009, Parmentier2013}. Intriguingly, models with TiO cold trapping can also fit the ensemble of WFC3 spectra \citep[see Figure 4 in][]{Mansfield2021}. This fact combined with the rarity of TiO detections suggests that cold-trapping of Ti might be a common phenomenon in hot and ultra-hot Jupiter atmospheres. More work is needed to explore this emerging population-level trend.

\citet{lothringer21} have proposed that the ratio of refractory and volatile species abundances is an important tracer of planet formation. KELT-20b is one of the few planets where Fe, H$_2$O, and CO spectral features are present in the thermal emission spectrum. The thermal emission spectrum typically arises from deeper atmospheric layers than the transmission spectrum, which can probe such high altitudes that mass fractionation of different chemical species can be an issue, thus complicating the interpretation of abundance measurements. By exhibiting Fe, H$_2$O, and CO features that arise from the bulk atmosphere, KELT-20b in principle presents an important opportunity to connect Fe/O and Fe/C abundance ratios to models of giant planet formation.

We have constrained the abundances of Fe\,\textsc{i}, H$_2$O, and CO for KELT-20b by performing a unified retrieval on ground-based, high-resolution optical spectroscopy and space-based low-resolution infrared spectroscopy. This extends our previous work by \citet{bro17}, which only conditioned an analysis of high-resolution data on a low-resolution retrieval, and the work of \citet{Brogi2019}, which developed our joint retrieval framework on simulated data. To our knowledge only one other such retrieval has been performed on real data before, by \citet{Gandhi19}.

Our measured Fe\,\textsc{i}, H$_2$O, and CO abundances and their ratios are consistent with a solar elemental Fe, C, and O under the assumption of 1D radiative-convective-thermochemical equilibrium. Further observations of KELT-20b could build on this work to determine more precise abundances of these elements as a constraint on the planet's formation. Of particular value would be a measurement of the mass of the planet, broader wavelength space-based spectroscopy from the \textit{James Webb Space Telescope}, and ground-based observations to probe the H$_2$O and CO lines at higher spectral resolution \citep[e.g.,][]{line2021, Pelletier2021, vansluijs22, yan22b, holmberg22}. In addition, advances in retrieval techniques are needed to ensure that the derived abundances are also accurate. A key limitation of the current state-of-the-art retrievals applied to high-resolution spectroscopy is the predominant assumption of 1D atmospheres and constant-with-altitude abundances. Future work that incorporates variations of temperature and abundance with longitude and altitude \citep[e.g.][]{gandhi22} will be critical to ensuring valid composition inferences as we continue to push towards higher fidelity datasets.

\begin{acknowledgements}
The University of Chicago group acknowledges funding for the MAROON-X project from the David and Lucile Packard Foundation, the Heising-Simons Foundation, the Gordon and Betty Moore Foundation, the Gemini Observatory, the NSF (award number 2108465), and NASA (grant number 80NSSC22K0117). We thank the staff of the Gemini Observatory for their assistance with the commissioning and operation of the instrument. J.L.B.\ and M.R.L.\ acknowledge support from NASA XRP grant 80NSSC19K0293. M.B.\ acknowledges support from the STFC research grant ST/T000406/1. This work was enabled by observations made from the Gemini North telescope, located within the Maunakea Science Reserve and adjacent to the summit of Maunakea. We are grateful for the privilege of observing the Universe from a place that is unique in both its astronomical quality and its cultural significance.
\end{acknowledgements}

\facilities{Gemini-North (MAROON-X), Hubble Space Telescope(WFC3), and Spitzer(IRAC)}
\software{\texttt{astropy} \citep{astropy:2018}, \texttt{barycorrpy} \citep{barycorrpy}, \texttt{corner} \citep{corner}, \texttt{matplotlib} \citep{matplotlib:2007}, \texttt{numpy} \citep{numpy:2020}, \texttt{python} \citep{python3:2009}, \texttt{scipy} \citep{2020SciPy-NMeth}}

\newpage

\appendix
\begin{figure}
\begin{center}
\includegraphics[height=\textheight]{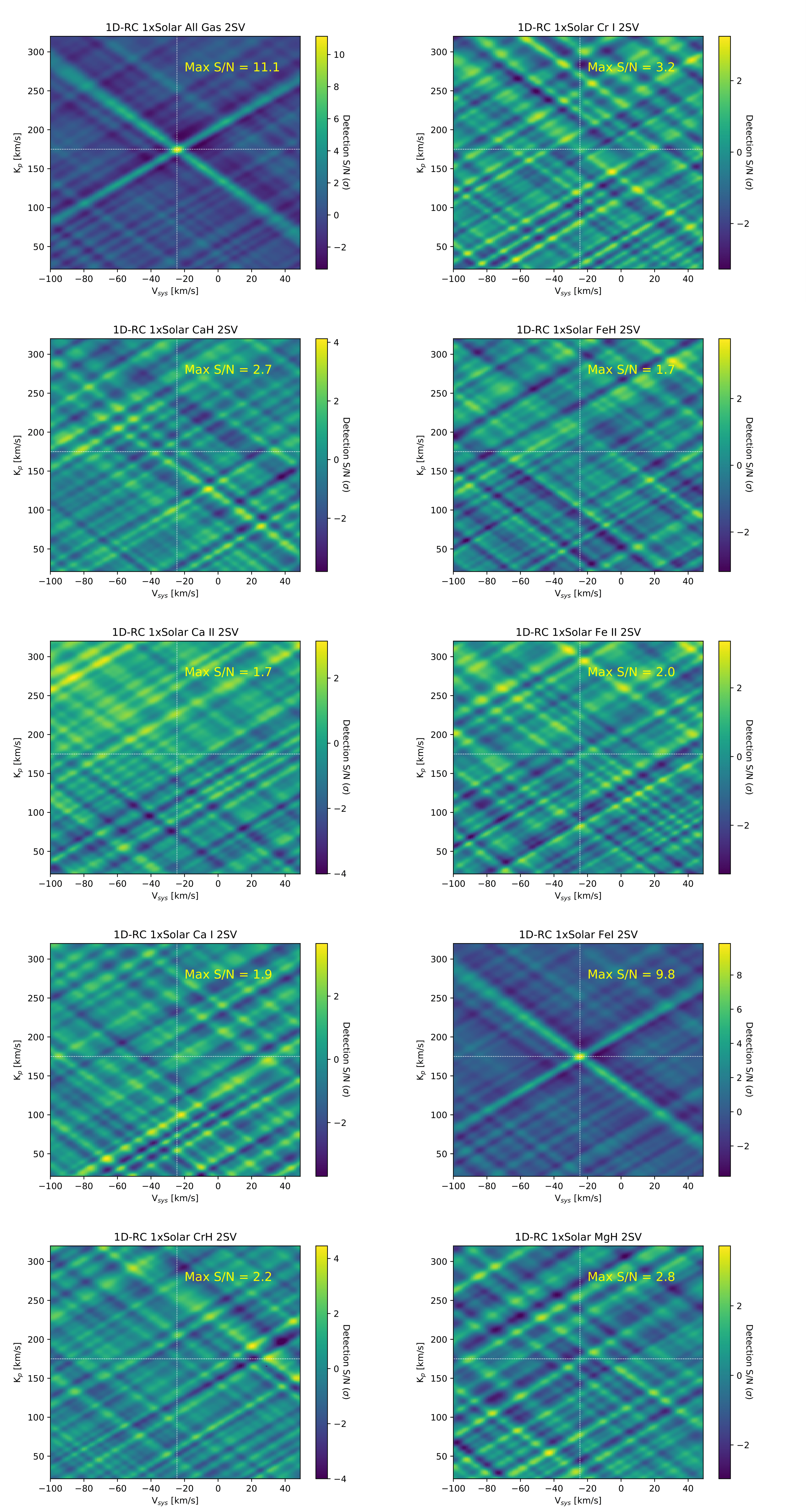}
\caption{Maps of the CCF in the planet velocity ($K_{p}$) vs.\ system velocity ($V_{\text{sys}}$) plane using the combined dataset and all 1D-RC model spectra explored (as labeled above each map). Dotted lines correspond to the known system velocity and exoplanet orbital speed of -24.5\,km\,s$^{-1}$ and 175\,km\,s$^{-1}$, respectively, as in \citet{borsa22}. Continued below.}
\label{fig:CCF_maps1}
\end{center}
\end{figure}

\begin{figure}
\begin{center}
\includegraphics[height=\textheight]{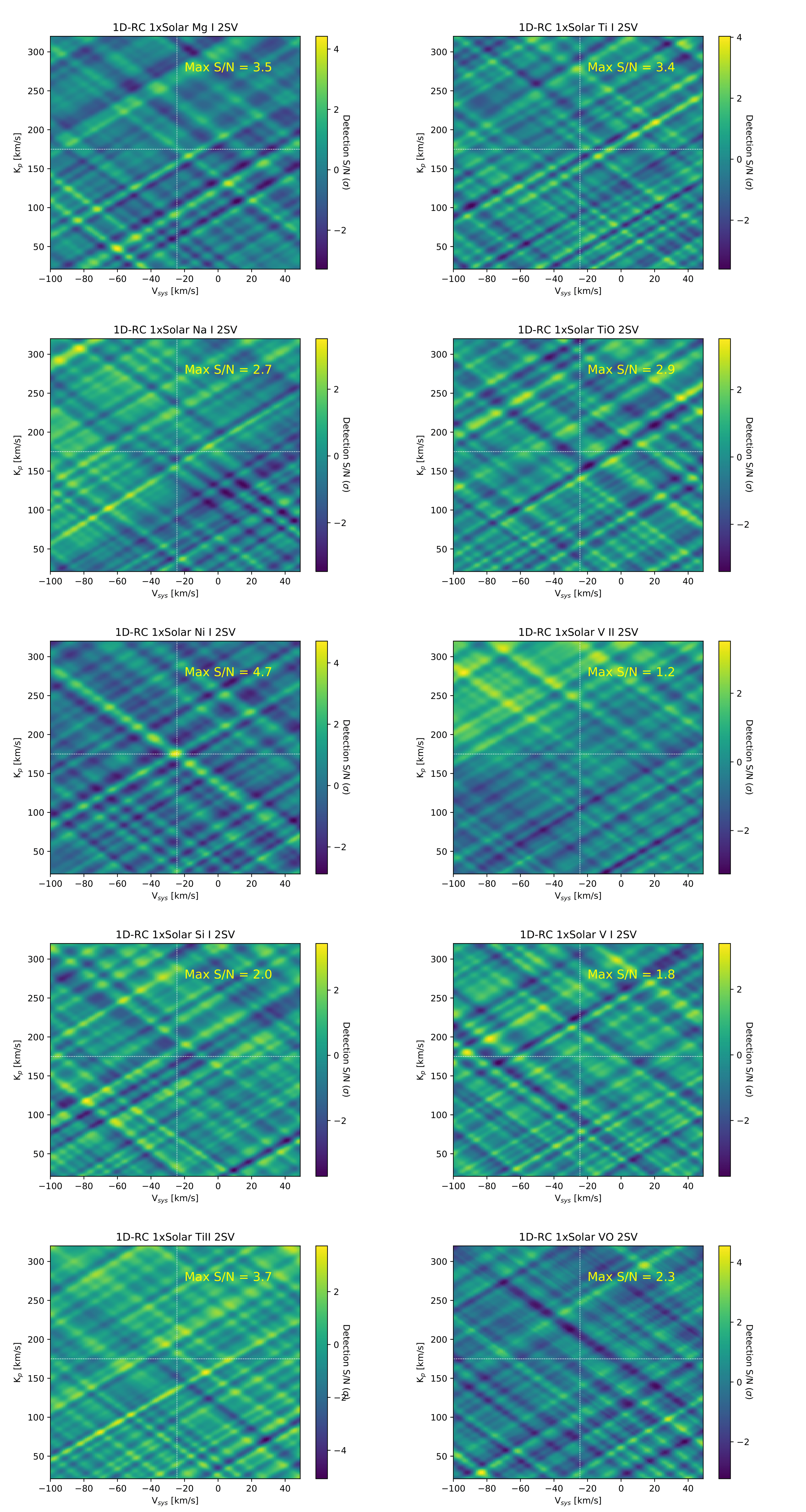}
\caption{Continuation of Fig \ref{fig:CCF_maps1}}
\label{fig:CCF_maps2}
\end{center}
\end{figure}

\begin{figure*}[h!]
\begin{center}
\includegraphics[width=\textwidth]{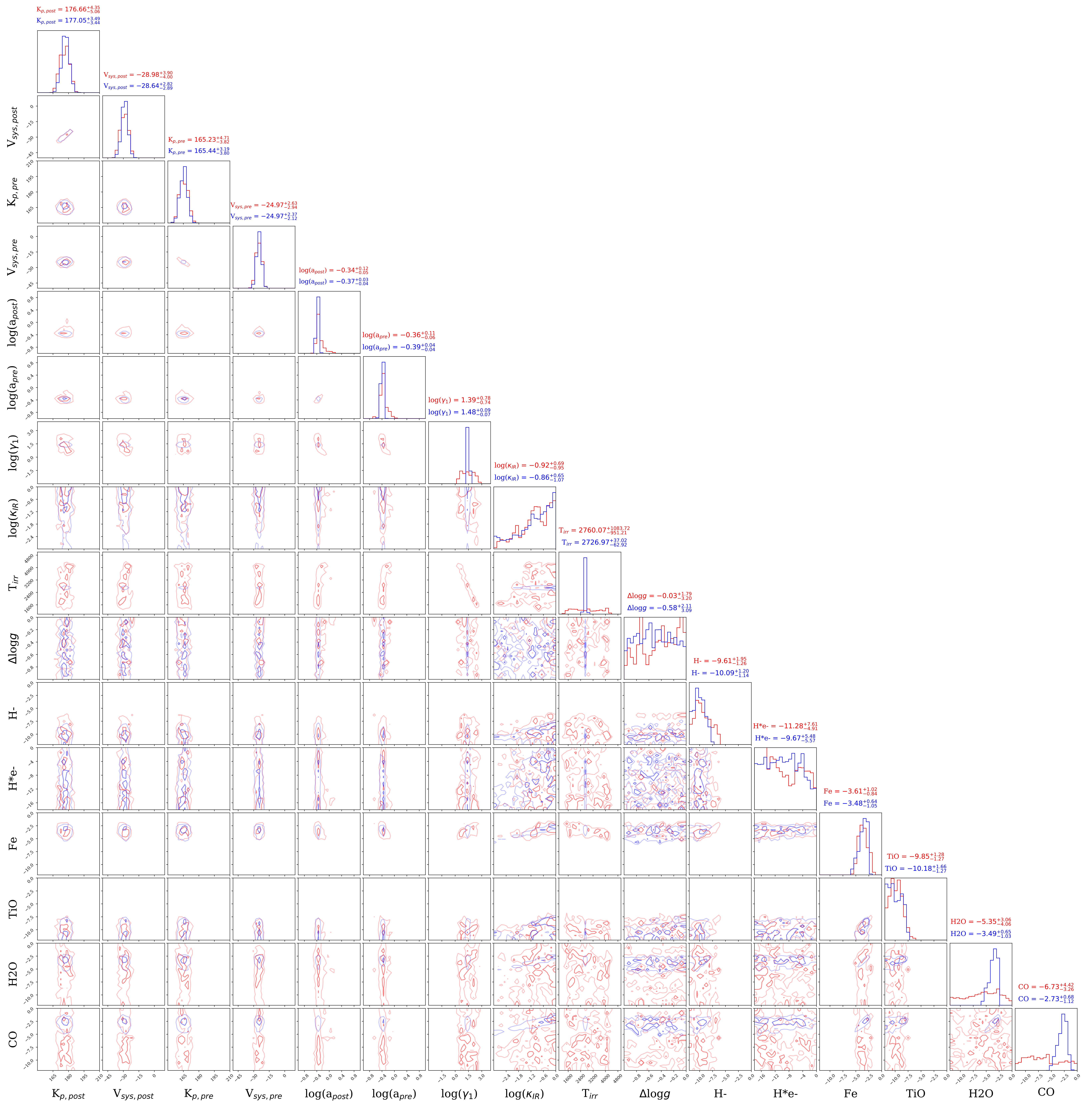}
\caption{Full retrieval results for KELT-20b. Red indicates the retrieval on just the MAROON-X data (both channels and both nights together), and blue indicates the joint retrieval with the \textit{HST}/WFC3 and \textit{Spitzer}/IRAC data. The numerical values above each histogram panel are the median and 68\% confidence interval. Both datasets by themselves provide sufficient evidence for the inversion. Isothermal atmospheres (i.e. log($\gamma_1$)=0) are strongly ruled out by the MAROON-X data alone as there is a hard edge to the log($\gamma_1$) in the MAROON-X Only data. There are no posterior samples below a log($\gamma_1$) value of 0.27 and this minimum value is excluded at $>$\,99.974\%.}
\label{fig:staircase}
\end{center}
\end{figure*}

\begin{figure*}[h!]
\begin{center}
\includegraphics[width=\textwidth]{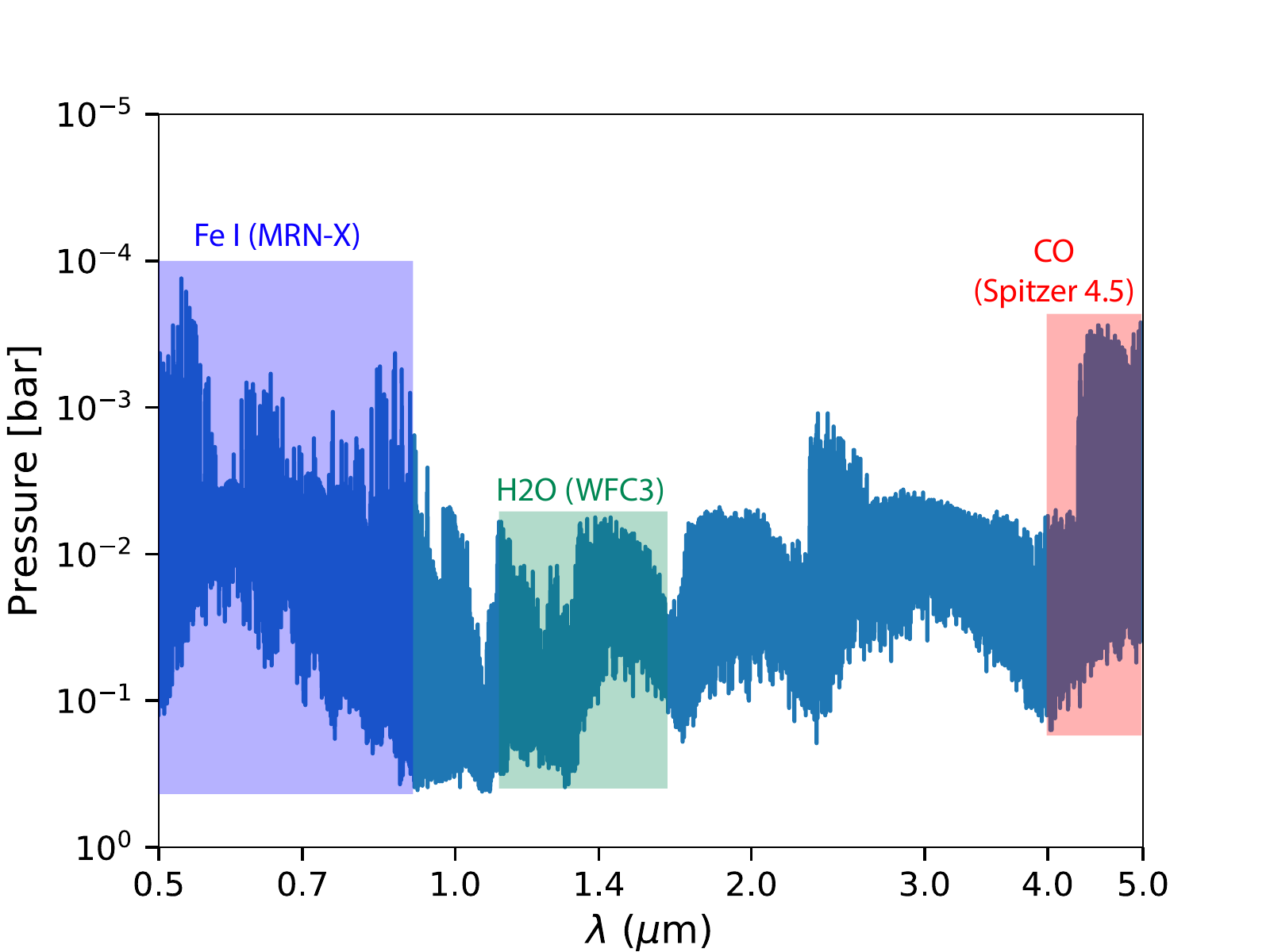}
\caption{Pressure levels probed by the observations as dictated by the location of the $\tau=2/3$ level at each wavelength. As described in the text, the retrieval can't precisely constrain the exact location of temperature with pressure due to the log($\kappa_{IR}$)-log$g$-metallicity degeneracy. Here, we show a representative surface from the 3$\times$solar 1D-RC model. The approximate pressure levels probed by each data-set by which species are indicated by the shaded boxes.   }
\label{fig:tau_sfc}
\end{center}
\end{figure*}

\bibliography{kelt20_dayside_2022}{}

\begin{thebibliography}{}
\expandafter\ifx\csname natexlab\endcsname\relax\def\natexlab#1{#1}\fi
\providecommand{\url}[1]{\href{#1}{#1}}
\providecommand{\dodoi}[1]{doi:~\href{http://doi.org/#1}{\nolinkurl{#1}}}
\providecommand{\doeprint}[1]{\href{http://ascl.net/#1}{\nolinkurl{http://ascl.net/#1}}}
\providecommand{\doarXiv}[1]{\href{https://arxiv.org/abs/#1}{\nolinkurl{https://arxiv.org/abs/#1}}}

\bibitem[{{Arcangeli} {et~al.}(2018){Arcangeli}, {D{\'e}sert}, {Line}, {Bean},
  {Parmentier}, {Stevenson}, {Kreidberg}, {Fortney}, {Mansfield}, \&
  {Showman}}]{Arcangeli2018}
{Arcangeli}, J., {D{\'e}sert}, J.-M., {Line}, M.~R., {et~al.} 2018, \apjl, 855,
  L30, \dodoi{10.3847/2041-8213/aab272}

\bibitem[{{Astropy Collaboration} {et~al.}(2018){Astropy Collaboration},
  {Price-Whelan}, {Sip{\H{o}}cz}, {G{\"u}nther}, {Lim}, {Crawford}, {Conseil},
  {Shupe}, {Craig}, {Dencheva}, {Ginsburg}, {Vand erPlas}, {Bradley},
  {P{\'e}rez-Su{\'a}rez}, {de Val-Borro}, {Aldcroft}, {Cruz}, {Robitaille},
  {Tollerud}, {Ardelean}, {Babej}, {Bach}, {Bachetti}, {Bakanov}, {Bamford},
  {Barentsen}, {Barmby}, {Baumbach}, {Berry}, {Biscani}, {Boquien}, {Bostroem},
  {Bouma}, {Brammer}, {Bray}, {Breytenbach}, {Buddelmeijer}, {Burke},
  {Calderone}, {Cano Rodr{\'\i}guez}, {Cara}, {Cardoso}, {Cheedella}, {Copin},
  {Corrales}, {Crichton}, {D'Avella}, {Deil}, {Depagne}, {Dietrich}, {Donath},
  {Droettboom}, {Earl}, {Erben}, {Fabbro}, {Ferreira}, {Finethy}, {Fox},
  {Garrison}, {Gibbons}, {Goldstein}, {Gommers}, {Greco}, {Greenfield},
  {Groener}, {Grollier}, {Hagen}, {Hirst}, {Homeier}, {Horton}, {Hosseinzadeh},
  {Hu}, {Hunkeler}, {Ivezi{\'c}}, {Jain}, {Jenness}, {Kanarek}, {Kendrew},
  {Kern}, {Kerzendorf}, {Khvalko}, {King}, {Kirkby}, {Kulkarni}, {Kumar},
  {Lee}, {Lenz}, {Littlefair}, {Ma}, {Macleod}, {Mastropietro}, {McCully},
  {Montagnac}, {Morris}, {Mueller}, {Mumford}, {Muna}, {Murphy}, {Nelson},
  {Nguyen}, {Ninan}, {N{\"o}the}, {Ogaz}, {Oh}, {Parejko}, {Parley}, {Pascual},
  {Patil}, {Patil}, {Plunkett}, {Prochaska}, {Rastogi}, {Reddy Janga},
  {Sabater}, {Sakurikar}, {Seifert}, {Sherbert}, {Sherwood-Taylor}, {Shih},
  {Sick}, {Silbiger}, {Singanamalla}, {Singer}, {Sladen}, {Sooley},
  {Sornarajah}, {Streicher}, {Teuben}, {Thomas}, {Tremblay}, {Turner},
  {Terr{\'o}n}, {van Kerkwijk}, {de la Vega}, {Watkins}, {Weaver}, {Whitmore},
  {Woillez}, {Zabalza}, \& {Astropy Contributors}}]{astropy:2018}
{Astropy Collaboration}, {Price-Whelan}, A.~M., {Sip{\H{o}}cz}, B.~M., {et~al.}
  2018, \aj, 156, 123, \dodoi{10.3847/1538-3881/aabc4f}

\bibitem[{{Barman} {et~al.}(2001){Barman}, {Hauschildt}, \&
  {Allard}}]{Barman2001}
{Barman}, T.~S., {Hauschildt}, P.~H., \& {Allard}, F. 2001, \apj, 556, 885,
  \dodoi{10.1086/321610}

\bibitem[{{Bello-Arufe} {et~al.}(2022){Bello-Arufe}, {Buchhave},
  {Mendon{\c{c}}a}, {Tronsgaard}, {Heng}, {Hoeijmakers}, \&
  {Mayo}}]{belloarufe22}
{Bello-Arufe}, A., {Buchhave}, L.~A., {Mendon{\c{c}}a}, J.~M., {et~al.} 2022,
  arXiv e-prints, arXiv:2203.04969.
\newblock \doarXiv{2203.04969}

\bibitem[{{Benneke}(2015)}]{Benneke2015}
{Benneke}, B. 2015, ArXiv e-prints: 1504.07655.
\newblock \doarXiv{1504.07655}

\bibitem[{Bernath(2020)}]{Bernath2020MOLLIST}
Bernath, P.~F. 2020, \jqsrt, 240, 106687,
  \dodoi{https://doi.org/10.1016/j.jqsrt.2019.106687}

\bibitem[{{Borsa} {et~al.}(2022){Borsa}, {Giacobbe}, {Bonomo}, {Brogi}, {Pino},
  {Fossati}, {Lanza}, {Nascimbeni}, {Sozzetti}, {Amadori}, {Benatti}, {Biazzo},
  {Bignamini}, {Boschin}, {Claudi}, {Cosentino}, {Covino}, {Desidera},
  {Fiorenzano}, {Guilluy}, {Harutyunyan}, {Maggio}, {Maldonado}, {Mancini},
  {Micela}, {Molinari}, {Molinaro}, {Pagano}, {Pedani}, {Piotto}, {Poretti},
  {Rainer}, {Scandariato}, \& {Stoev}}]{borsa22}
{Borsa}, F., {Giacobbe}, P., {Bonomo}, A.~S., {et~al.} 2022, arXiv e-prints,
  arXiv:2204.04948.
\newblock \doarXiv{2204.04948}

\bibitem[{{Brogi} {et~al.}(2017){Brogi}, {Line}, {Bean}, {D{\'e}sert}, \&
  {Schwarz}}]{bro17}
{Brogi}, M., {Line}, M., {Bean}, J., {D{\'e}sert}, J.-M., \& {Schwarz}, H.
  2017, \apjl, 839, L2, \dodoi{10.3847/2041-8213/aa6933}

\bibitem[{Brogi \& Line(2019)}]{Brogi2019}
Brogi, M., \& Line, M.~R. 2019, AJ, 157, 114

\bibitem[{{Casasayas-Barris} {et~al.}(2019){Casasayas-Barris}, {Pall{\'e}},
  {Yan}, {Chen}, {Kohl}, {Stangret}, {Parviainen}, {Helling}, {Watanabe},
  {Czesla}, {Fukui}, {Monta{\~n}{\'e}s-Rodr{\'\i}guez}, {Nagel}, {Narita},
  {Nortmann}, {Nowak}, {Schmitt}, \& {Zapatero Osorio}}]{Casasayas19}
{Casasayas-Barris}, N., {Pall{\'e}}, E., {Yan}, F., {et~al.} 2019, \aap, 628,
  A9, \dodoi{10.1051/0004-6361/201935623}

\bibitem[{Foreman-Mackey(2016)}]{corner}
Foreman-Mackey, D. 2016, The Journal of Open Source Software, 1, 24,
  \dodoi{10.21105/joss.00024}

\bibitem[{{Fortney} {et~al.}(2008){Fortney}, {Lodders}, {Marley}, \&
  {Freedman}}]{Fortney2008}
{Fortney}, J.~J., {Lodders}, K., {Marley}, M.~S., \& {Freedman}, R.~S. 2008,
  \apj, 678, 1419, \dodoi{10.1086/528370}

\bibitem[{{Fu} {et~al.}(2022){Fu}, {Sing}, {Lothringer}, {Deming}, {Ih},
  {Kempton}, {Malik}, {Komacek}, {Mansfield}, \& {Bean}}]{Fu22}
{Fu}, G., {Sing}, D.~K., {Lothringer}, J.~D., {et~al.} 2022, \apjl, 925, L3,
  \dodoi{10.3847/2041-8213/ac4968}

\bibitem[{{Gandhi} {et~al.}(2022){Gandhi}, {Kesseli}, {Snellen}, {Brogi},
  {Wardenier}, {Parmentier}, {Welbanks}, \& {Savel}}]{gandhi22}
{Gandhi}, S., {Kesseli}, A., {Snellen}, I., {et~al.} 2022, \mnras,
  \dodoi{10.1093/mnras/stac1744}

\bibitem[{{Gandhi} {et~al.}(2019){Gandhi}, {Madhusudhan}, {Hawker}, \&
  {Piette}}]{Gandhi19}
{Gandhi}, S., {Madhusudhan}, N., {Hawker}, G., \& {Piette}, A. 2019, \aj, 158,
  228, \dodoi{10.3847/1538-3881/ab4efc}

\bibitem[{{Gharib-Nezhad} {et~al.}(2021){Gharib-Nezhad}, {Iyer}, {Line},
  {Freedman}, {Marley}, \& {Batalha}}]{GharibNezhad21}
{Gharib-Nezhad}, E., {Iyer}, A.~R., {Line}, M.~R., {et~al.} 2021, \apjs, 254,
  34, \dodoi{10.3847/1538-4365/abf504}

\bibitem[{{GharibNezhad} {et~al.}(2013){GharibNezhad}, {Shayesteh}, \&
  {Bernath}}]{GharibNezhad2013}
{GharibNezhad}, E., {Shayesteh}, A., \& {Bernath}, P.~F. 2013, \mnras, 432,
  2043, \dodoi{10.1093/mnras/stt510}

\bibitem[{{Giacobbe} {et~al.}(2021){Giacobbe}, {Brogi}, {Gandhi}, {Cubillos},
  {Bonomo}, {Sozzetti}, {Fossati}, {Guilluy}, {Carleo}, {Rainer},
  {Harutyunyan}, {Borsa}, {Pino}, {Nascimbeni}, {Benatti}, {Biazzo},
  {Bignamini}, {Chubb}, {Claudi}, {Cosentino}, {Covino}, {Damasso}, {Desidera},
  {Fiorenzano}, {Ghedina}, {Lanza}, {Leto}, {Maggio}, {Malavolta}, {Maldonado},
  {Micela}, {Molinari}, {Pagano}, {Pedani}, {Piotto}, {Poretti}, {Scandariato},
  {Yurchenko}, {Fantinel}, {Galli}, {Lodi}, {Sanna}, \& {Tozzi}}]{Giacobbe2021}
{Giacobbe}, P., {Brogi}, M., {Gandhi}, S., {et~al.} 2021, \nat, 592, 205,
  \dodoi{10.1038/s41586-021-03381-x}

\bibitem[{{Grimm} \& {Heng}(2015)}]{Grimm2015}
{Grimm}, S.~L., \& {Heng}, K. 2015, \apj, 808, 182,
  \dodoi{10.1088/0004-637X/808/2/182}

\bibitem[{{Grimm} {et~al.}(2021){Grimm}, {Malik}, {Kitzmann},
  {Guzm{\'a}n-Mesa}, {Hoeijmakers}, {Fisher}, {Mendon{\c{c}}a}, {Yurchenko},
  {Tennyson}, {Alesina}, {Buchschacher}, {Burnier}, {Segransan}, {Kurucz}, \&
  {Heng}}]{Grimm2021}
{Grimm}, S.~L., {Malik}, M., {Kitzmann}, D., {et~al.} 2021, \apjs, 253, 30,
  \dodoi{10.3847/1538-4365/abd773}

\bibitem[{{Guillot}(2010)}]{Guillot2010}
{Guillot}, T. 2010, \aap, 520, A27+, \dodoi{10.1051/0004-6361/200913396}

\bibitem[{Harris {et~al.}(2020)Harris, Millman, van~der Walt, Gommers,
  Virtanen, Cournapeau, Wieser, Taylor, Berg, Smith, Kern, Picus, Hoyer, van
  Kerkwijk, Brett, Haldane, del R{\'{i}}o, Wiebe, Peterson,
  G{\'{e}}rard-Marchant, Sheppard, Reddy, Weckesser, Abbasi, Gohlke, \&
  Oliphant}]{numpy:2020}
Harris, C.~R., Millman, K.~J., van~der Walt, S.~J., {et~al.} 2020, Nature, 585,
  357, \dodoi{10.1038/s41586-020-2649-2}

\bibitem[{{Hauschildt} {et~al.}(1999){Hauschildt}, {Allard}, \&
  {Baron}}]{hauschildt:1999}
{Hauschildt}, P.~H., {Allard}, F., \& {Baron}, E. 1999, \apj, 512, 377,
  \dodoi{10.1086/306745}

\bibitem[{{Hoeijmakers} {et~al.}(2020){Hoeijmakers}, {Cabot}, {Zhao},
  {Buchhave}, {Tronsgaard}, {Davis}, {Kitzmann}, {Grimm}, {Cegla}, {Bourrier},
  {Ehrenreich}, {Heng}, {Lovis}, \& {Fischer}}]{hoeijmakers20}
{Hoeijmakers}, H.~J., {Cabot}, S. H.~C., {Zhao}, L., {et~al.} 2020, \aap, 641,
  A120, \dodoi{10.1051/0004-6361/202037437}

\bibitem[{{Holmberg} \& {Madhusudhan}(2022)}]{holmberg22}
{Holmberg}, M., \& {Madhusudhan}, N. 2022, arXiv e-prints, arXiv:2206.10621.
\newblock \doarXiv{2206.10621}

\bibitem[{{Hubeny} {et~al.}(2003){Hubeny}, {Burrows}, \&
  {Sudarsky}}]{Hubeny2003}
{Hubeny}, I., {Burrows}, A., \& {Sudarsky}, D. 2003, \apj, 594, 1011,
  \dodoi{10.1086/377080}

\bibitem[{Hunter(2007)}]{matplotlib:2007}
Hunter, J.~D. 2007, Computing in Science \& Engineering, 9, 90,
  \dodoi{10.1109/MCSE.2007.55}

\bibitem[{{John}(1988)}]{John1988}
{John}, T.~L. 1988, \aap, 193, 189

\bibitem[{{Johnson} {et~al.}(2022){Johnson}, {Wang}, {Pai Asnodkar}, {Bonomo},
  {Gaudi}, {Henning}, {Ilyin}, {Keles}, {Malavolta}, {Mallonn},
  {Molaverdikhani}, {Nascimbeni}, {Patience}, {Poppenhaeger}, {Scandariato},
  {Schlawin}, {Shkolnik}, {Sicilia}, {Sozzetti}, {Strassmeier}, {Veillet}, \&
  {Yan}}]{Johnson22}
{Johnson}, M.~C., {Wang}, J., {Pai Asnodkar}, A., {et~al.} 2022, arXiv
  e-prints, arXiv:2205.12162.
\newblock \doarXiv{2205.12162}

\bibitem[{Kanodia \& Wright(2018)}]{barycorrpy}
Kanodia, S., \& Wright, J. 2018, Research Notes of the {AAS}, 2, 4,
  \dodoi{10.3847/2515-5172/aaa4b7}

\bibitem[{{Kasper} {et~al.}(2021){Kasper}, {Bean}, {Line}, {Seifahrt},
  {St{\"u}rmer}, {Pino}, {D{\'e}sert}, \& {Brogi}}]{kasper2021}
{Kasper}, D., {Bean}, J.~L., {Line}, M.~R., {et~al.} 2021, \apjl, 921, L18,
  \dodoi{10.3847/2041-8213/ac30e1}

\bibitem[{{Kesseli} {et~al.}(2020){Kesseli}, {Snellen}, {Alonso-Floriano},
  {Molli{\`e}re}, \& {Serindag}}]{Kesseli20}
{Kesseli}, A.~Y., {Snellen}, I.~A.~G., {Alonso-Floriano}, F.~J.,
  {Molli{\`e}re}, P., \& {Serindag}, D.~B. 2020, \aj, 160, 228,
  \dodoi{10.3847/1538-3881/abb59c}

\bibitem[{{Kreidberg} {et~al.}(2014){Kreidberg}, {Bean}, {D{\'e}sert},
  {Benneke}, {Deming}, {Stevenson}, {Seager}, {Berta-Thompson}, {Seifahrt}, \&
  {Homeier}}]{kreidberg14}
{Kreidberg}, L., {Bean}, J.~L., {D{\'e}sert}, J.-M., {et~al.} 2014, \nat, 505,
  69, \dodoi{10.1038/nature12888}

\bibitem[{{Langeveld} {et~al.}(2022){Langeveld}, {Madhusudhan}, \&
  {Cabot}}]{Langeveld22}
{Langeveld}, A.~B., {Madhusudhan}, N., \& {Cabot}, S. H.~C. 2022, arXiv
  e-prints, arXiv:2205.01623.
\newblock \doarXiv{2205.01623}

\bibitem[{{Line} {et~al.}(2021){Line}, {Brogi}, {Bean}, {Gandhi}, {Zalesky},
  {Parmentier}, {Smith}, {Mace}, {Mansfield}, {Kempton}, {Fortney}, {Shkolnik},
  {Patience}, {Rauscher}, {D{\'e}sert}, \& {Wardenier}}]{line2021}
{Line}, M.~R., {Brogi}, M., {Bean}, J.~L., {et~al.} 2021, \nat, 598, 580,
  \dodoi{10.1038/s41586-021-03912-6}

\bibitem[{{Lothringer} \& {Barman}(2019)}]{lothringer19}
{Lothringer}, J.~D., \& {Barman}, T. 2019, \apj, 876, 69,
  \dodoi{10.3847/1538-4357/ab1485}

\bibitem[{{Lothringer} {et~al.}(2018){Lothringer}, {Barman}, \&
  {Koskinen}}]{lothringer18}
{Lothringer}, J.~D., {Barman}, T., \& {Koskinen}, T. 2018, \apj, 866, 27,
  \dodoi{10.3847/1538-4357/aadd9e}

\bibitem[{{Lothringer} {et~al.}(2021){Lothringer}, {Rustamkulov}, {Sing},
  {Gibson}, {Wilson}, \& {Schlaufman}}]{lothringer21}
{Lothringer}, J.~D., {Rustamkulov}, Z., {Sing}, D.~K., {et~al.} 2021, \apj,
  914, 12, \dodoi{10.3847/1538-4357/abf8a9}

\bibitem[{{Lund} {et~al.}(2017){Lund}, {Rodriguez}, {Zhou}, {Gaudi}, {Stassun},
  {Johnson}, {Bieryla}, {Oelkers}, {Stevens}, {Collins}, {Penev}, {Quinn},
  {Latham}, {Villanueva}, {Eastman}, {Kielkopf}, {Oberst}, {Jensen}, {Cohen},
  {Joner}, {Stephens}, {Relles}, {Corfini}, {Gregorio}, {Zambelli}, {Esquerdo},
  {Calkins}, {Berlind}, {Ciardi}, {Dressing}, {Patel}, {Gagnon}, {Gonzales},
  {Beatty}, {Siverd}, {Labadie-Bartz}, {Kuhn}, {Col{\'o}n}, {James}, {Pepper},
  {Fulton}, {McLeod}, {Stockdale}, {Calchi Novati}, {DePoy}, {Gould},
  {Marshall}, {Trueblood}, {Trueblood}, {Johnson}, {Wright}, {McCrady},
  {Wittenmyer}, {Johnson}, {Sergi}, {Wilson}, \& {Sliski}}]{lund17}
{Lund}, M.~B., {Rodriguez}, J.~E., {Zhou}, G., {et~al.} 2017, \aj, 154, 194,
  \dodoi{10.3847/1538-3881/aa8f95}

\bibitem[{{Mansfield} {et~al.}(2021){Mansfield}, {Line}, {Bean}, {Fortney},
  {Parmentier}, {Wiser}, {Kempton}, {Gharib-Nezhad}, {Sing},
  {L{\'o}pez-Morales}, {Baxter}, {D{\'e}sert}, {Swain}, \&
  {Roudier}}]{Mansfield2021}
{Mansfield}, M., {Line}, M.~R., {Bean}, J.~L., {et~al.} 2021, Nature Astronomy,
  5, 1224, \dodoi{10.1038/s41550-021-01455-4}

\bibitem[{{McKemmish} {et~al.}(2019){McKemmish}, {Masseron}, {Hoeijmakers},
  {P{\'e}rez-Mesa}, {Grimm}, {Yurchenko}, \& {Tennyson}}]{mckemmish19}
{McKemmish}, L.~K., {Masseron}, T., {Hoeijmakers}, H.~J., {et~al.} 2019,
  \mnras, 488, 2836, \dodoi{10.1093/mnras/stz1818}

\bibitem[{{McKemmish} {et~al.}(2016){McKemmish}, {Yurchenko}, \&
  {Tennyson}}]{Mckemmish16}
{McKemmish}, L.~K., {Yurchenko}, S.~N., \& {Tennyson}, J. 2016, Molecular
  Physics, 114, 3232, \dodoi{10.1080/00268976.2016.1225994}

\bibitem[{{Nugroho} {et~al.}(2020){Nugroho}, {Gibson}, {de Mooij}, {Watson},
  {Kawahara}, \& {Merritt}}]{nugroho20}
{Nugroho}, S.~K., {Gibson}, N.~P., {de Mooij}, E. J.~W., {et~al.} 2020, \mnras,
  496, 504, \dodoi{10.1093/mnras/staa1459}

\bibitem[{{Parmentier} {et~al.}(2013){Parmentier}, {Showman}, \&
  {Lian}}]{Parmentier2013}
{Parmentier}, V., {Showman}, A.~P., \& {Lian}, Y. 2013, \aap, 558, A91,
  \dodoi{10.1051/0004-6361/201321132}

\bibitem[{{Pelletier} {et~al.}(2021){Pelletier}, {Benneke}, {Darveau-Bernier},
  {Boucher}, {Cook}, {Piaulet}, {Coulombe}, {Artigau}, {Lafreni{\`e}re},
  {Deslile}, {Allart}, {Doyon}, {Donati}, {Fouqu{\'e}}, {Moutou}, {Cadieux},
  {Delfosse}, {H{\'e}brard}, {Martins}, {Martioli}, \&
  {Vandal}}]{Pelletier2021}
{Pelletier}, S., {Benneke}, B., {Darveau-Bernier}, A., {et~al.} 2021, arXiv
  e-prints, arXiv:2105.10513.
\newblock \doarXiv{2105.10513}

\bibitem[{{Pino} {et~al.}(2020){Pino}, {D{\'e}sert}, {Brogi}, {Malavolta},
  {Wyttenbach}, {Line}, {Hoeijmakers}, {Fossati}, {Bonomo}, {Nascimbeni},
  {Panwar}, {Affer}, {Benatti}, {Biazzo}, {Bignamini}, {Borsa}, {Carleo},
  {Claudi}, {Cosentino}, {Covino}, {Damasso}, {Desidera}, {Giacobbe},
  {Harutyunyan}, {Lanza}, {Leto}, {Maggio}, {Maldonado}, {Mancini}, {Micela},
  {Molinari}, {Pagano}, {Piotto}, {Poretti}, {Rainer}, {Scandariato},
  {Sozzetti}, {Allart}, {Borsato}, {Bruno}, {Di Fabrizio}, {Ehrenreich},
  {Fiorenzano}, {Frustagli}, {Lavie}, {Lovis}, {Magazz{\`u}}, {Nardiello},
  {Pedani}, \& {Smareglia}}]{pino20}
{Pino}, L., {D{\'e}sert}, J.-M., {Brogi}, M., {et~al.} 2020, \apjl, 894, L27,
  \dodoi{10.3847/2041-8213/ab8c44}

\bibitem[{{Piskorz} {et~al.}(2018){Piskorz}, {Buzard}, {Line}, {Knutson},
  {Benneke}, {Crockett}, {Lockwood}, {Blake}, {Barman}, {Bender}, {Deming}, \&
  {Johnson}}]{pis18}
{Piskorz}, D., {Buzard}, C., {Line}, M.~R., {et~al.} 2018, \aj, 156, 133,
  \dodoi{10.3847/1538-3881/aad781}

\bibitem[{{Polyansky} {et~al.}(2018){Polyansky}, {Kyuberis}, {Zobov},
  {Tennyson}, {Yurchenko}, \& {Lodi}}]{Polyansky2018}
{Polyansky}, O.~L., {Kyuberis}, A.~A., {Zobov}, N.~F., {et~al.} 2018, \mnras,
  480, 2597, \dodoi{10.1093/mnras/sty1877}

\bibitem[{{Prinoth} {et~al.}(2022){Prinoth}, {Hoeijmakers}, {Kitzmann},
  {Sandvik}, {Seidel}, {Lendl}, {Borsato}, {Thorsbro}, {Anderson}, {Barrado},
  {Kravchenko}, {Allart}, {Bourrier}, {Cegla}, {Ehrenreich}, {Fisher}, {Lovis},
  {Guzm{\'a}n-Mesa}, {Grimm}, {Hooton}, {Morris}, {Oreshenko}, {Pino}, \&
  {Heng}}]{prinoth22}
{Prinoth}, B., {Hoeijmakers}, H.~J., {Kitzmann}, D., {et~al.} 2022, Nature
  Astronomy, 6, 449, \dodoi{10.1038/s41550-021-01581-z}

\bibitem[{{Rainer} {et~al.}(2021){Rainer}, {Borsa}, {Pino}, {Frustagli},
  {Brogi}, {Biazzo}, {Bonomo}, {Carleo}, {Claudi}, {Gratton}, {Lanza},
  {Maggio}, {Maldonado}, {Mancini}, {Micela}, {Scandariato}, {Sozzetti},
  {Buchschacher}, {Cosentino}, {Covino}, {Ghedina}, {Gonzalez}, {Leto}, {Lodi},
  {Martinez Fiorenzano}, {Molinari}, {Molinaro}, {Nardiello}, {Oliva},
  {Pagano}, {Pedani}, {Piotto}, \& {Poretti}}]{Rainer21}
{Rainer}, M., {Borsa}, F., {Pino}, L., {et~al.} 2021, \aap, 649, A29,
  \dodoi{10.1051/0004-6361/202039247}

\bibitem[{{Seifahrt} {et~al.}(2016){Seifahrt}, {Bean}, {St{\"u}rmer}, {Gers},
  {Grobler}, {Reed}, \& {Jones}}]{seifahrt16}
{Seifahrt}, A., {Bean}, J.~L., {St{\"u}rmer}, J., {et~al.} 2016, in Society of
  Photo-Optical Instrumentation Engineers (SPIE) Conference Series, Vol. 9908,
  Ground-based and Airborne Instrumentation for Astronomy VI, ed. C.~J.
  {Evans}, L.~{Simard}, \& H.~{Takami}, 990818, \dodoi{10.1117/12.2232069}

\bibitem[{{Seifahrt} {et~al.}(2018){Seifahrt}, {St{\"u}rmer}, {Bean}, \&
  {Schwab}}]{seifahrt18}
{Seifahrt}, A., {St{\"u}rmer}, J., {Bean}, J.~L., \& {Schwab}, C. 2018, in
  Society of Photo-Optical Instrumentation Engineers (SPIE) Conference Series,
  Vol. 10702, Ground-based and Airborne Instrumentation for Astronomy VII, ed.
  C.~J. {Evans}, L.~{Simard}, \& H.~{Takami}, 107026D,
  \dodoi{10.1117/12.2312936}

\bibitem[{{Seifahrt} {et~al.}(2020){Seifahrt}, {Bean}, {St{\"u}rmer}, {Kasper},
  {Gers}, {Schwab}, {Zechmeister}, {Stef{\'a}nsson}, {Montet}, {Dos Santos},
  {Peck}, {White}, \& {Tapia}}]{seifahrt20}
{Seifahrt}, A., {Bean}, J.~L., {St{\"u}rmer}, J., {et~al.} 2020, in Society of
  Photo-Optical Instrumentation Engineers (SPIE) Conference Series, Vol. 11447,
  Society of Photo-Optical Instrumentation Engineers (SPIE) Conference Series,
  114471F, \dodoi{10.1117/12.2561564}

\bibitem[{{Sheppard} {et~al.}(2017){Sheppard}, {Mandell}, {Tamburo}, {Gandhi},
  {Pinhas}, {Madhusudhan}, \& {Deming}}]{Sheppard2017}
{Sheppard}, K.~B., {Mandell}, A.~M., {Tamburo}, P., {et~al.} 2017, \apjl, 850,
  L32, \dodoi{10.3847/2041-8213/aa9ae9}

\bibitem[{{Spiegel} {et~al.}(2009){Spiegel}, {Silverio}, \&
  {Burrows}}]{Spiegel2009}
{Spiegel}, D.~S., {Silverio}, K., \& {Burrows}, A. 2009, \apj, 699, 1487,
  \dodoi{10.1088/0004-637X/699/2/1487}

\bibitem[{{Stangret} {et~al.}(2020){Stangret}, {Casasayas-Barris}, {Pall{\'e}},
  {Yan}, {S{\'a}nchez-L{\'o}pez}, \& {L{\'o}pez-Puertas}}]{Stangret20}
{Stangret}, M., {Casasayas-Barris}, N., {Pall{\'e}}, E., {et~al.} 2020, \aap,
  638, A26, \dodoi{10.1051/0004-6361/202037541}

\bibitem[{{Su{\'a}rez Mascare{\~n}o} {et~al.}(2020){Su{\'a}rez Mascare{\~n}o},
  {Faria}, {Figueira}, {Lovis}, {Damasso}, {Gonz{\'a}lez Hern{\'a}ndez},
  {Rebolo}, {Cristiani}, {Pepe}, {Santos}, {Zapatero Osorio}, {Adibekyan},
  {Hojjatpanah}, {Sozzetti}, {Murgas}, {Abreu}, {Affolter}, {Alibert},
  {Aliverti}, {Allart}, {Allende Prieto}, {Alves}, {Amate}, {Avila}, {Baldini},
  {Bandi}, {Barros}, {Bianco}, {Benz}, {Bouchy}, {Broeng}, {Cabral},
  {Calderone}, {Cirami}, {Coelho}, {Conconi}, {Coretti}, {Cumani}, {Cupani},
  {D'Odorico}, {Deiries}, {Delabre}, {Di Marcantonio}, {Dumusque},
  {Ehrenreich}, {Fragoso}, {Genolet}, {Genoni}, {G{\'e}nova Santos}, {Hughes},
  {Iwert}, {Kerber}, {Knusdstrup}, {Landoni}, {Lavie}, {Lillo-Box}, {Lizon},
  {Lo Curto}, {Maire}, {Manescau}, {Martins}, {M{\'e}gevand}, {Mehner},
  {Micela}, {Modigliani}, {Molaro}, {Monteiro}, {Monteiro}, {Moschetti},
  {Mueller}, {Nunes}, {Oggioni}, {Oliveira}, {Pall{\'e}}, {Pariani},
  {Pasquini}, {Poretti}, {Rasilla}, {Redaelli}, {Riva}, {Santana Tschudi},
  {Santin}, {Santos}, {Segovia}, {Sosnowska}, {Sousa}, {Span{\`o}}, {Tenegi},
  {Udry}, {Zanutta}, \& {Zerbi}}]{suarez20}
{Su{\'a}rez Mascare{\~n}o}, A., {Faria}, J.~P., {Figueira}, P., {et~al.} 2020,
  \aap, 639, A77, \dodoi{10.1051/0004-6361/202037745}

\bibitem[{{Talens} {et~al.}(2018){Talens}, {Justesen}, {Albrecht}, {McCormac},
  {Van Eylen}, {Otten}, {Murgas}, {Palle}, {Pollacco}, {Stuik}, {Spronck},
  {Lesage}, {Grundahl}, {Fredslund Andersen}, {Antoci}, \&
  {Snellen}}]{Talens18}
{Talens}, G.~J.~J., {Justesen}, A.~B., {Albrecht}, S., {et~al.} 2018, \aap,
  612, A57, \dodoi{10.1051/0004-6361/201731512}

\bibitem[{Van~Rossum \& Drake(2009)}]{python3:2009}
Van~Rossum, G., \& Drake, F.~L. 2009, Python 3 Reference Manual (Scotts Valley,
  CA: CreateSpace)

\bibitem[{{van Sluijs} {et~al.}(2022){van Sluijs}, {Birkby}, {Lothringer},
  {Lee}, {Crossfield}, {Parmentier}, {Brogi}, {Kulesa}, {McCarthy}, {Powell},
  \& {Charbonneau}}]{vansluijs22}
{van Sluijs}, L., {Birkby}, J.~L., {Lothringer}, J., {et~al.} 2022, arXiv
  e-prints, arXiv:2203.13234.
\newblock \doarXiv{2203.13234}

\bibitem[{Virtanen {et~al.}(2020)Virtanen, Gommers, Oliphant, Haberland, Reddy,
  Cournapeau, Burovski, Peterson, Weckesser, Bright, {van der Walt}, Brett,
  Wilson, Millman, Mayorov, Nelson, Jones, Kern, Larson, Carey, Polat, Feng,
  Moore, {VanderPlas}, Laxalde, Perktold, Cimrman, Henriksen, Quintero, Harris,
  Archibald, Ribeiro, Pedregosa, {van Mulbregt}, \& {SciPy 1.0
  Contributors}}]{2020SciPy-NMeth}
Virtanen, P., Gommers, R., Oliphant, T.~E., {et~al.} 2020, Nature Methods, 17,
  261, \dodoi{10.1038/s41592-019-0686-2}

\bibitem[{{Yan} {et~al.}(2022{\natexlab{a}}){Yan}, {Reiners}, {Pall{\'e}},
  {Shulyak}, {Stangret}, {Molaverdikhani}, {Nortmann}, {Molli{\`e}re},
  {Henning}, {Casasayas-Barris}, {Cont}, {Chen}, {Czesla},
  {S{\'a}nchez-L{\'o}pez}, {L{\'o}pez-Puertas}, {Ribas}, {Quirrenbach},
  {Caballero}, {Amado}, {Galad{\'\i}-Enr{\'\i}quez}, {Khalafinejad}, {Lara},
  {Montes}, {Morello}, {Nagel}, {Sedaghati}, {Zapatero Osorio}, \&
  {Zechmeister}}]{Yan22}
{Yan}, F., {Reiners}, A., {Pall{\'e}}, E., {et~al.} 2022{\natexlab{a}}, \aap,
  659, A7, \dodoi{10.1051/0004-6361/202142395}

\bibitem[{{Yan} {et~al.}(2022{\natexlab{b}}){Yan}, {Pall{\'e}}, {Reiners},
  {Casasayas-Barris}, {Cont}, {Stangret}, {Nortmann}, {Molli{\`e}re},
  {Henning}, {Chen}, \& {Molaverdikhani}}]{yan22b}
{Yan}, F., {Pall{\'e}}, E., {Reiners}, A., {et~al.} 2022{\natexlab{b}}, \aap,
  661, L6, \dodoi{10.1051/0004-6361/202243503}

\end{thebibliography}
\bibliographystyle{aasjournal}

\end{document}